\documentclass{elsart}
\usepackage{epsfig}
\usepackage{amssymb}

\begin{document}
\begin{frontmatter}


\title{Spin observables in deuteron-proton radiative capture at intermediate energies}

\author[kvi,ira]{A.A.~Mehmandoost-Khajeh-Dad},
\author[kvi]{H.R.~Amir-Ahmadi},
\author[kvi]{J.C.S.~Bacelar},
\author[kvi]{A.M.~van~den~Berg},
\author[kvi]{R.~Castelijns},
\author[han]{A.~Deltuva\thanksref{label1}},
\author[kvi]{E.D.~van~Garderen},
\author[boc]{W.~Gl\"ockle},
\author[jag]{J.~Golak},
\author[kvi]{N.~Kalantar-Nayestanaki},
\author[jap]{H.~Kamada},
\author[kvi]{M.~Ki\v{s}\thanksref{label2}},
\author[ira]{R.~Koohi-Fayegh-Dehkordi},
\author[kvi]{H.~L\"ohner},
\author[kvi]{M.~Mahjour-Shafiei\thanksref{label3}},
\author[kvi]{H.~Mardanpur},
\author[kvi]{J.G.~Messchendorp},
\author[usa]{A.~Nogga},
\author[han]{P.~Sauer},
\author[kvi]{S.V.~Shende},
\author[jag]{R.~Skibinski},
\author[jag]{H.~Wita\l{a}},
\author[kvi]{H.J.~W\"ortche}

\thanks[label1]{Present address: Centro de Fisica Nuclear da Universidade de Lisboa, Lisboa, Portugal}
\thanks[label2]{Present address: Rudjer Boskovic Institute, Zagreb, Croatia}
\thanks[label3]{Present address: Tehran University, Tehran, Iran}

\address[kvi]{Kernfysisch Versneller Instituut (KVI), Groningen, The Netherlands}
\address[ira]{Mashad Ferdowsi University, Mashad, Iran}
\address[han]{Institute for Theoretical Physics, University of Hannover, Hannover, Germany}
\address[boc]{Instit\"ut f\"ur Theoretische Physik II, Ruhr-Universit\"at Bochum, Bochum, Germany}
\address[jag]{M. Smoluchowski Institute of Physics, Jagiellonian University, Krak\'ow, Poland}
\address[jap]{Department of Physics, Faculty of Engineering, Kyushu Institute of Technology, 1-1 Sensuicho, Tobata, Kitakyushu 804-8550, Japan}
\address[usa]{Institute for Nuclear Theory, University of Washington, Seattle, USA}




\maketitle

\begin{abstract}
A radiative deuteron-proton capture experiment was carried out at KVI 
using polarized-deuteron beams at incident energies of 55,
66.5, and 90~MeV/nucleon. Vector and tensor-analyzing 
powers were obtained for a large angular range.
The results are interpreted with the help of Faddeev calculations, which are 
based on modern two- and three-nucleon potentials. Our data are described well 
by the calculations, and disagree significantly with the observed tensor 
anomaly at RCNP.
\end{abstract}

\end{frontmatter}

The structure of nuclei and the dynamics of reactions are
described by the strong nuclear force governing the 
nucleon-nucleon interaction~\cite{Bro96}.
The longest-range two-nucleon force (2NF) is due
to the exchange of a pion~\cite{Sto93}, an idea
that goes back to the work of Yukawa in 1935. At
present, 2NF models exist which provide an excellent
description of the high-quality data base of proton-proton
and neutron-proton scattering and of the properties
of the deuteron. For heavier nuclei, Green's function 
Monte-Carlo calculations employing 2NFs clearly underestimate 
the experimental binding energies~\cite{Wir02}, and therefore show that
2NF are not sufficient to describe the three-nucleon
system and heavier systems accurately. 

In the last decades, our understanding of the
three-nucleon system has improved significantly.
High-precision data at intermediate energies in $Nd$ elastic 
scattering~\cite{Erm01,Sek02,Erm03} for a large energy interval 
together with rigorous Faddeev calculations~\cite{Glo96} for 
the three-nucleon system have put large constraints on 
phenomenological three-nucleon forces (3NF). These studies 
are supported by calculations based
on $\chi$PT at lower energies, which are expected to provide model-independent 
predictions for the complete structure of the 
3NF~\cite{Ord92,Kol94,Epe98,Fri99} in the near future. 

The radiative deuteron-proton capture reaction, $p+d$$\rightarrow$$^3$He+$\gamma$,
is an interesting channel
since it involves a large momentum transfer and therefore probes 
high-momentum components of the wave functions involved in the matrix element.
In addition, the coupling with a photon makes this reaction 
sensitive to electromagnetic currents involved in the three-nucleon system.
These aspects make the radiative capture process a unique tool to extend 
the above described three-nucleon force studies. 

In the last few years, the interest in
the radiative $Nd$ capture channel has increased. This is partly due to the
presently available theoretical techniques which solve the three-nucleon system
rigorously. In contrast to the elastic $Nd$ scattering data, however, 
the experimental data base on radiative $Nd$ 
capture is much poorer. In particular, in the intermediate energy 
range ($\sim$50-200~MeV/nucleon), below the pion-production threshold, the available 
data~\cite{Cam84,Pic87,Joh98,Mes00} are scarce and in general lack precision 
or completeness in angular coverage and the number of observables. 

Recently, a precision deuteron-proton radiative capture experiment\cite{Sag03} 
at RCNP was conducted using a vector and tensor-polarized deuteron beam impinging
on a proton target at an incident deuteron energy of 100~MeV/nucleon.
Interestingly, the preliminary results on tensor-analyzing powers~\cite{Sag03} 
show large discrepancies with present-day calculations. These deviations
were found to be larger than a factor three for $A_{xx}$ in comparison
with several different model approaches. As a result, the authors 
speculated about possible existence of new forces or new mechanisms that 
are sensitive to tensor observables. A confirmation of this intriguing
tensor anomaly in $pd$-radiative capture is clearly needed. In addition,
a study of the energy dependence is necessary in order to understand the 
origin of these discrepancies.

In this paper, deuteron-proton radiative capture data on vector and
tensor-analyzing powers obtained at KVI are presented along with a comparison with
two theoretical approaches. The first calculation by the Bochum-Cracow
group~\cite{Ski03,Gol00} is a Faddeev calculation with the AV18 2NF and
an additional phenomenological Urbana IX 3NF as input. The coupling
with a photon is described via two different approaches. The first
approach supplements the single-nucleon current operator by 
exchange currents which take explicitly into account $\pi$- and $\rho$-like
meson-exchange contributions. Alternatively,
the meson-exchange currents are included using the extended 
Siegert theorem. In this form, electric and magnetic multipoles are kept
to very high orders for the one-body operator. As a consequence of the Siegert approximation,
only many-body currents in the electric multipoles are accounted for.
The second calculation is from the Hannover theory group~\cite{Del04},
which describes the process using the purely nucleonic charge-dependent CD-Bonn
potential and its coupled-channel extension CD-Bonn+$\Delta$. Within this
approach, the $\Delta$-isobar excitation mediates an effective 3NF
with prominent Fujita-Miyazawa and Illinois ring type 
contributions. These contributions are based on the exchange of
$\pi$, $\rho$, $\omega$, and $\sigma$ mesons and are mutually consistent. 
The electromagnetic current in the
Hannover approach has one-baryon and two-baryon contributions and couples
to nucleonic and $\Delta$-isobar channels. Therefore, the $\Delta$-isobar
generates consistently effective two- and three-nucleon currents in addition 
to a 3NF.

\begin{figure}[t]
\centerline{\includegraphics[height=0.5\textwidth]{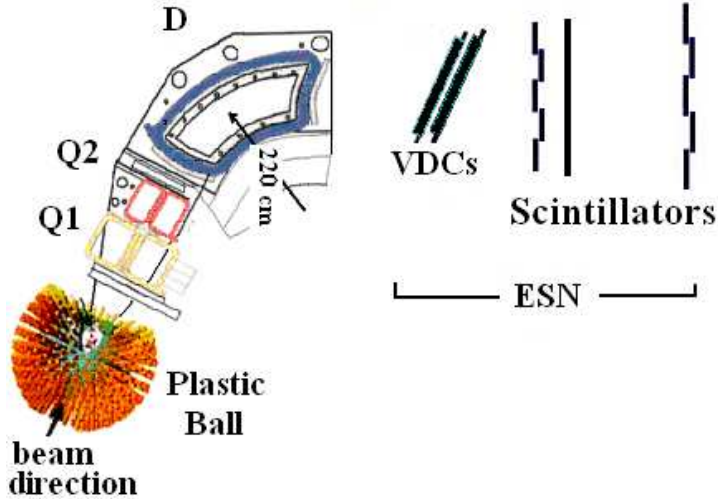}}
\caption{A sketch of the experimental setup at KVI, which measured the 
deuteron-proton
radiative capture reaction. The Big-Bite spectrometer (BBS)
was employed to detect the $^3$He at forward angles. The photon was detected between 50$^\circ$ 
and {160}$^\circ$ using the Plastic Ball detector. A cryogenic proton target
was placed at the center of the Plastic Ball.}
\label{fig:setup}
\end{figure}

\begin{figure}[t]
\centerline{\includegraphics[height=0.6\textwidth]{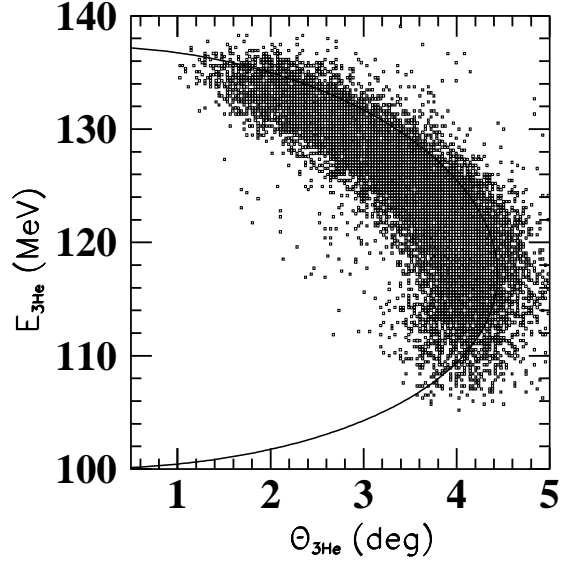}}
\caption{The measured energy of $^3$He versus its scattering angle measured by 
the BBS placed at a scattering angle of {3.5}$^\circ$. The data are obtained with a 90~MeV/nucleon
deuteron beam. The solid line represents the expected kinematical dependence of the
deuteron-proton radiative capture process. The width of the band is due to energy straggling
in the target.}
\label{fig:locie}
\end{figure}

The experiment was carried out in autumn of 2003 at KVI, 
The Netherlands. Beams of vector and tensor-polarized deuterons were
produced in an atomic-beam-type ion source (POLIS) and accelerated with the
superconducting cyclotron, AGOR, up to 55, 66.5, and 90~MeV/nucleon.
The beam with an intensity of $\approx$0.5~nA impinged on a 4.5~mm thick liquid-hydrogen
target. The $^3$He particle and the photon were detected using a coincidence setup
between the Big-Bite Spectrometer (BBS) and the Plastic Ball detector (PB), respectively. 
The setup is depicted in Fig.~\ref{fig:setup}. The magnetic spectrometer 
BBS~\cite{Ber95}, with an angular acceptance of $\approx${3.8}$^\circ$, with its Euro-Supernova focal-plane
detection system (ESN)~\cite{Wor01} was placed at various angles between {1.7}$^\circ$ and
{3.5}$^\circ$  for different energies to cover as large a center-of-mass angular range as possible.
With this detector, nearly the complete $^3$He phase space is covered.
The energy and angle resolutions were dominated by straggling in the target.
However, the two-body reconstruction was enough to identify the events 
adequately (see Fig.~\ref{fig:locie}). The PB detector~\cite{Bad82}
was equipped with $\approx$500 $\Delta$E-E
phoswich modules covering photon scattering angles between 50$^\circ$ and 160$^\circ$ with
complete azimuthal acceptance. Each module contains a 4~mm thick CaF$_2$ layer (slow component) 
glued on to a 356~mm thick scintillator (fast component), which allows to discriminate 
photons, leptons, and protons from each other.
The PB detector measures the scattering angle 
of the photon with a resolution of {6}$^\circ$ and with an efficiency of $\approx$50\% .

Figure~\ref{fig:locie} demonstrates the quality of the $^3$He analysis with the BBS 
detection system. Here, measured energy of the $^3$He is plotted against 
its scattering angle obtained from the BBS for an incident deuteron beam 
energy of 90~MeV/nucleon. The 
solid line represents the expected kinematical correlation for the radiative capture 
reaction. A coincidence with a photon detected by the PB is required in this 
plot. The data coincide with the expected kinematical correlation. Background channels, like
radiative break-up, will not follow such a behaviour. This demonstrates that the reaction of
interest can be identified unambigously.

Vector and tensor-analyzing powers in the deuteron-proton radiative capture process
were obtained by employing a beam of polarized deuterons. Five different polarization
states were provided by the ion source with theoretical polarization values of
(p$_Z$,p$_{ZZ}$)=
(0,0),(2/3,0), \- (--2/3,0),(0,1),(0,--2).
 In this notation, 
p$_Z$ and p$_{ZZ}$ represent the vector and tensor polarizations of the deuteron ion beam at the source. 
The beam polarization with a typical value of 70-80\% of the theoretical value was monitored 
regularly using the In-Beam Polarimeter (IBP)~\cite{Bie01}. 


In this paper, we present a measurement of vector ($A_y(d)$) and tensor ($A_{yy}$, $A_{zz}$)
analyzing powers of the $\vec d+p\rightarrow {\rm ^3He}+\gamma$ reaction. These observables
were extracted by making use of the dependence on the azimuthal angle, $\phi$, of the reaction
rate, $I(\theta,\phi)$, according to~\cite{Ohl72}
\begin{eqnarray}
\frac{I(\theta,\phi)}{I_0(\theta)}= 1 &+& \frac{3}{2}{\rm p}_{Z} A_y(\theta)\cos\phi \nonumber \\
 &-& \frac{1}{2}{\rm p}_{ZZ} A_{zz}(\theta)\sin^2\phi\\\
 &+& \frac{1}{2}{\rm p}_{ZZ} A_{yy}(\theta)\cos 2\phi,\nonumber
\end{eqnarray}
\noindent where $I_0(\theta)$ is the reaction rate for an unpolarized beam and $\theta$ is 
the polar angle of the $\gamma-p$ system in the center-of-mass.
Note that at KVI the polarization vector is perpendicular to the beam direction.
Exploiting the above equation, the vector analyzing power, $A_y(d)$, is obtained from
the reaction rates for ion-source states (p$_Z$,p$_{ZZ}$)=(2/3,0) ($N^+$) and (--2/3,0) ($N^-$)
integrating $\phi$ from --{55}$^\circ$ to {55}$^\circ$ for a beam energy of 90~MeV/nucleon and
from --{65}$^\circ$ to {65}$^\circ$ for beam energies of 55 and 66.5~MeV/nucleon,
according to
\begin{eqnarray}
A_y(d)=-\frac{2}{3}\,\frac{N^+-N^-}{{\rm p}_Z^- N^+ - {\rm p}_Z^+ N^-},
\end{eqnarray} 
\noindent where p$_Z^+$ is the measured vector polarization for\- the (2/3,0) spin mode 
and p$_Z^-$ is the polarization for the (--2/3,0) spin mode. The azimuthal angle, $\phi$, 
is obtained from the PB. Corrections due to variations in the photon-detection efficiency are
properly taken into account by measuring these distribution using the unpolarized 
data of the radiative-capture reaction. Similarly, the tensor-analyzing power, $A_{zz}$, is
deduced using the states (p$_Z$,p$_{ZZ}$)=(0,1) and (0,--2) integrating $\phi$ from 35$^\circ$ 
to 55$^\circ$ (and from --35$^\circ$ to --55$^\circ$) for a beam energy of 90~MeV/nucleon and
integrating $\phi$ from 25$^\circ$ to 65$^\circ$ (and from --25$^\circ$ to --65$^\circ$) for 
a beam energy of 66.5~MeV/nucleon. According to Eq.~1, this integration should cancel 
the contribution of $A_{yy}$ to the cross section. Remaining contributions of $A_{yy}$ due to 
variations in the photon-detection efficiency in $\phi$ are properly corrected for. 
The tensor-analyzing power, $A_{yy}$, is obtained by integrating, for the same ion states, 
over the azimuthal angle $\phi$ from --15$^\circ$ to 15$^\circ$. Also here, non-vanishing 
contributions of the term
$\frac{1}{2}{\rm p}_{ZZ}A_{zz}\sin^2\phi$ in Eq.~1 due to a finite $\phi$ integration are 
small ($\sim$2\%) and corrected for.

\begin{figure}[h]
\centerline{\includegraphics[height=0.8\textwidth]{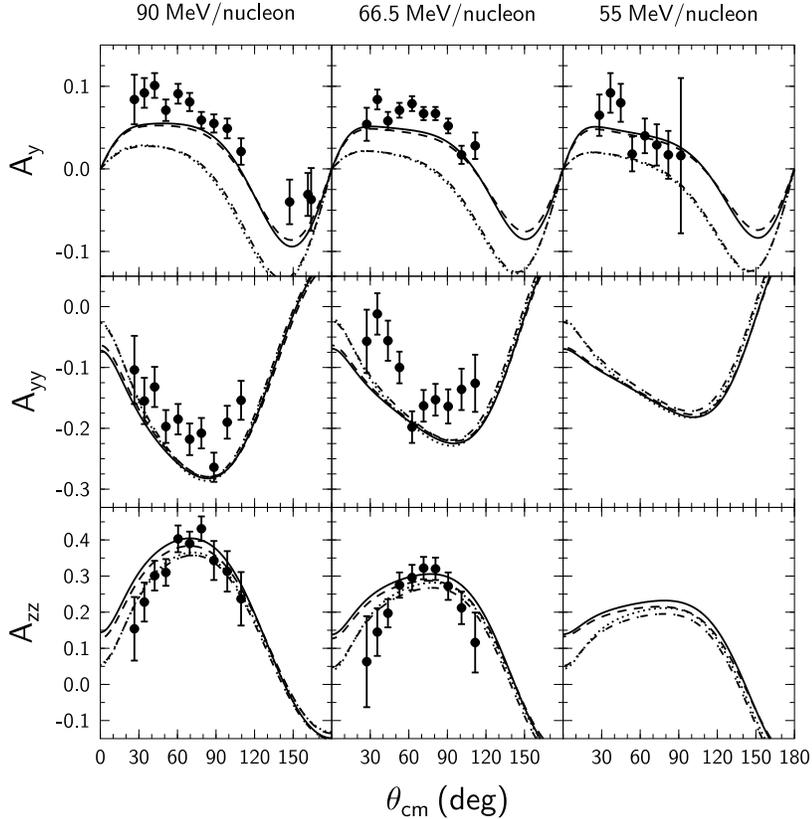}}
\caption{Polarization data for the deuteron-proton radiative capture reaction are compared to
Faddeev calculations by the Bochum-Cracow theory group. The data are shown as filled circles
with the statistical uncertainty indicated by error bar. The dotted and dot-dashed lines 
(which can hardly be distinguished) represent 
the results of the calculation using the Siegert approximation with the AV18 2NF as input and with 
the additional inclusion of the Urbana-IX 3NF, respectively. The dashed (2NF) and solid (2NF+3NF) 
lines are similar calculations for which meson-exchange currents are calculated using explicit 
$\pi$ and $\rho$ exchanges.}
\label{fig:results1}
\end{figure}

Figure~\ref{fig:results1} shows the results of the
deuteron-proton radiative capture experiment in comparison with
the calculation by the Bochum-Cracow group~\cite{Ski03,Gol00}.
Data for $A_y(d)$, $A_{yy}$, and $A_{zz}$ are presented as a function
of the $\gamma$-$p$ center-of-mass angle for three different
incident deuteron energies of 55, 66.5, and 90~MeV/nucleon.
Only statistical uncertainties are indicated by the error bars.
The systematic uncertainty due to the error in the beam polarization
is estimated to be less than 6\%.
The dotted lines are the result of the Faddeev calculation for which
the AV18 2NF is used as input. Meson-exchange currents are included
using the Siegert approximation. The dot-dashed lines
correspond to the same model and including the Urbana-IX 3NF.
An explicit inclusion of $\pi$ and $\rho$ like meson-exchange contributions
are represented by the dashed and solid lines. For the solid lines, the Urbana-IX
3NF was included, whereas the dashed lines only take into account the
AV18 2NF. The data for $A_y(d)$ clearly disagree with the calculation
in which meson-exchange contributions are constructed using the
Siegert approximation. This might point to large magnetic contributions which
are not properly included for many-body currents. The approximation with an explicit 
inclusion of $\pi$ and $\rho$ exchange agrees well with our data.
The effect of the 3NF is small within the framework of this approximation. At present,
no experiment will have sufficient sensitivity to prefer this calculation with
or without the inclusion of a 3NF.

\begin{figure}[h]
\centerline{\includegraphics[height=0.8\textwidth]{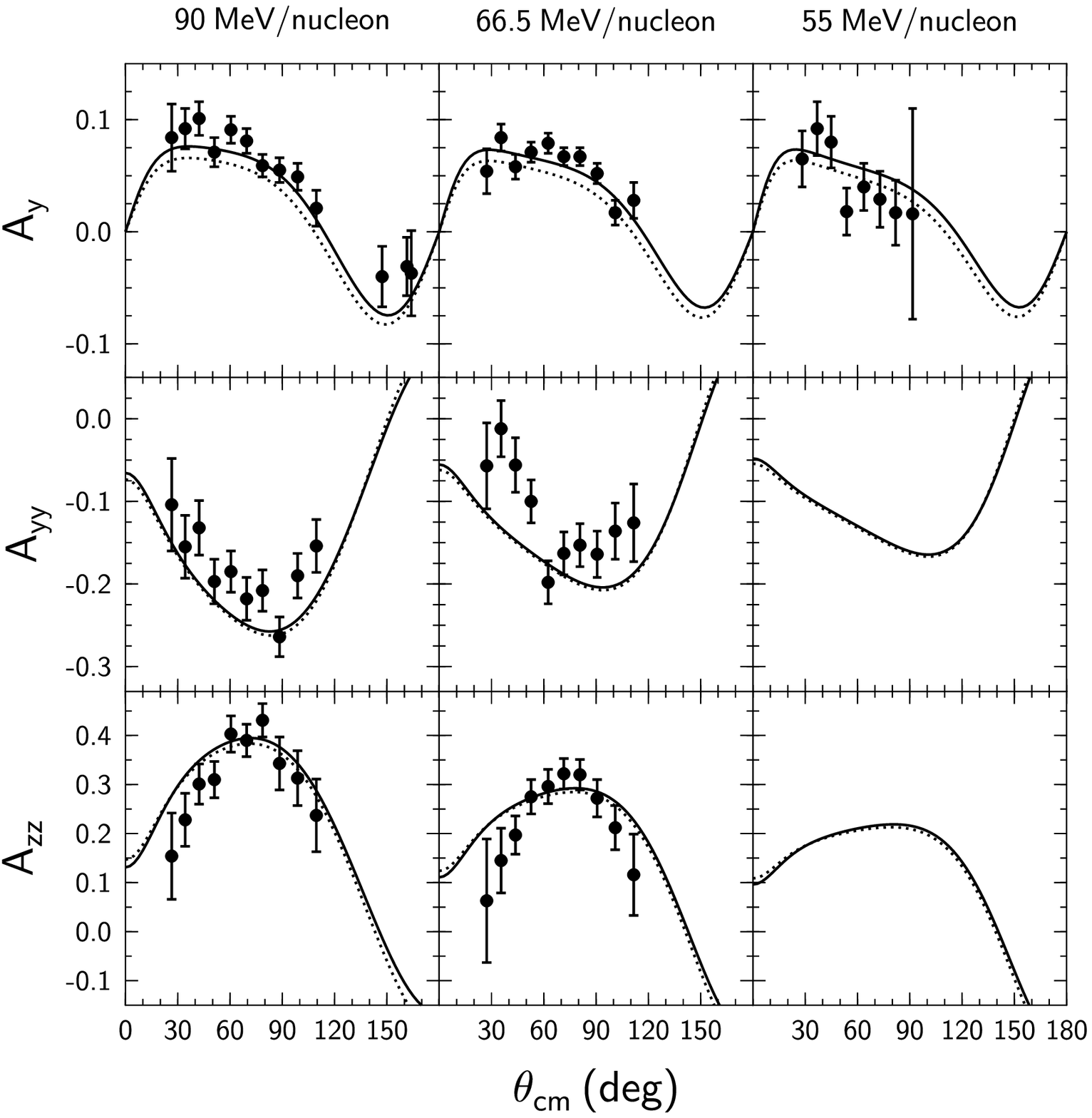}}
\caption{The same data as shown in Fig.~\ref{fig:results1} are compared with 
the predictions of the Hannover group. The dashed line represents the
calculation based on the CD-Bonn potential, whereas the solid line includes in addition 
contributions from the $\Delta$ isobar.}
\label{fig:results2}
\end{figure}

Figure~\ref{fig:results2} compares the same data as in Fig.~\ref{fig:results1} with the
predictions of the Hannover group~\cite{Del04}, based on the
purely nucleonic CD-Bonn potential and its coupled-channel extrension,
CD Bonn + $\Delta$, allowing for a single excitation of a nucleon
to a $\Delta$ isobar. The $\Delta$ mediates 3NFs and generates 
effective two- and three-nucleon currents in addition to irreducible 
one- and two-baryon contributions as described in detail in Ref.~\cite{Del04}. 
Note that the Hannover
calculation agrees reasonably well with our data for all polarization observables and all
energies. The effect of the $\Delta$ isobar is small at these energies and for these
observables.

The tensor-analyzing powers $A_{zz}$ and $A_{yy}$ hardly depend on the 
choice of approximation. The predictions from the Bochum-Cracow group are
similar to those of the Hannover group for these observables. Since the
tensor-analyzing power $A_{xx}$ is related to $A_{yy}$ and $A_{zz}$, 
via $A_{xx}$+$A_{yy}$+$A_{zz}$=0, we conclude that also this observable 
is rather well predicted by all models including MECs. Surprisingly,
a recent experiment conducted at RCNP
with a 100~MeV/nucleon incident deuteron beam showed huge deviations for $A_{xx}$ in
comparison with similar model predictions\cite{Sag03}. Our data taken at an
energy of 90~MeV/nucleon clearly do not show such large discrepancies, and therefore 
contradict the preliminary data of RCNP. Also at lower energies, no anomaly is
observed for the tensor-analyzing powers in the deuteron-proton radiative capture
process.

In summary, this paper presents data on vector and tensor-analyzing powers
in the deuteron-proton radiative capture process. The data were taken at
KVI with the Big-Bite Spectrometer and the almost-4$\pi$ Plastic Ball detection systems 
which measure the momentum vectors of the $^3$He and $\gamma$, respectively. 
A polarized deuteron beam was employed at
incident energies of 55, 66.5, and 90~MeV/nucleon. The reaction rate 
dependence on the 
azimuthal angle together with different combinations for the polarization states (p$_Z$, p$_{ZZ}$) 
were used to extract $A_y(d)$, $A_{yy}$, and $A_{zz}$ for a large angular range.
The results are interpreted using Faddeev calculations by the Bochum-Cracow group and 
by the Hannover group.
In general, our results agree reasonably well with predictions by the Hannover group and the predictions 
by the Bochum-Cracow group in case an explicit $\pi$ and $\rho$ exchange is used.
The large discrepancy of our data with the calculation by the Bochum-Cracow group using the Siegert 
approximation demonstrates the sensitivity to the treatment of the electromagnetic currents
in the radiative-capture reaction. In particular, the vector-analyzing powers are rather
sensitive to MECs making them a good testing ground for the details of exchange currents.
The calculations and experimental results presented in this
paper indicate that the effect of a phenomenological three-nucleon force and the contribution
of the $\Delta$ isobar are, at these energies, small. Therefore, these
observables are ideally suited to test electromagnetic currents and form factors in
a three-nucleon system. The study presented in this paper was partly motivated by a 
recent observation of a large
discrepancy between measured tensor-analyzing powers taken at RCNP and predictions by various
modern calculations. Our data taken at a similar energy do not show such a discrepancy and therefore
disagree with the preliminary RCNP data. 

The authors acknowledge the work by the cyclotron and ion-source groups at KVI for
delivering the high-quality beam used in these measurements and thank Muhsin Harakeh
for a careful proofreading of this Letter. 
This work is part of the research program of the 
``Stichting voor Fundamenteel Onderzoek der Materie'' (FOM) with financial 
support from the ``Nederlandse Organisatie voor Wetenschappelijk Onderzoek'' (NWO).
The work of the Cracow-Bochum group was supported by the Polish Committee for Scientific
Research under Grants No. 2P03B0825. The numerical calculations have been performed
on the Cray SV1 of the NIC in J\"ulich, Germany.


\end{document}